\documentclass[namedreferences,hyperref,optionalrh]{spr-sola}

\usepackage{graphicx}
\usepackage{color}
\usepackage{url}
\ifx \doiurl    \undefined \def \doiurl#1{\href{http://dx.doi.org/#1}{\textsf{DOI}}}\fi
\ifx \adsurl    \undefined \def \adsurl#1{\href{http://adsabs.harvard.edu/abs/#1}{\textsf{ADS}}}\fi

\newcommand{\aap}{    {\it Astron. Astrophys.}}

\newcommand{\apj}{    {\it Astrophys. J.}}

\newcommand{\pasj}{   {\it Pub. Astron. Soc. Japan}}

\newcommand{\solphys}{{\it Solar Phys.}}

\newcommand{\raa}{    {\it Res. Astron. Astrophys.}}

\begin{document}

\begin{article}
\begin{opening}

   \title{Correction for the Weakening Magnetic Field within the Sunspot Umbra Observed by ASO-S/FMG}

   \author[addressref={aff1,aff2},corref,email={xhq@bao.ac.cn}]{\inits{Haiqing}\fnm{Haiqing}~\snm{Xu}\orcid{0000-0003-4244-1077}}
   \author[addressref={aff1,aff2,aff3},corref,email={sjt@bao.ac.cn}]{\inits{Jiangtao}\fnm{Jiangtao}~\snm{Su}\orcid{0000-0002-5152-7318}}
   \author[addressref={aff1,aff2,aff3}]{\fnm{Suo}~\lnm{Liu}\orcid{0000-0002-1396-7603}}
   \author[addressref={aff1,aff2,aff3}]{\fnm{Yuanyong}~\lnm{Deng}\orcid{0000-0003-1988-4574}} 
   \author[addressref={aff1,aff2,aff3}]{\fnm{Xianyong}~\lnm{Bai}\orcid{0000-0003-2686-9153}}
   \author[addressref={aff1,aff2}]{\fnm{Jie}~\lnm{Chen}\orcid{0000-0001-7472-5539}}
   \author[addressref={aff1,aff2}]{\fnm{Xiaofan}~\lnm{Wang}\orcid{0000-0003-2776-3895}}
   \author[addressref={aff1,aff2}]{\fnm{Xiao}~\lnm{Yang}\orcid{0000-0003-1675-1995}}
   \author[addressref={aff1,aff2}]{\fnm{Yongliang}~\lnm{Song}\orcid{0000-0002-9961-4357}}

  \address[id=aff1]{National Astronomical Observatories, Chinese Academy of Sciences, Beijing, 100101, China}
  \address[id=aff2]{Key Laboratory of Solar Activity and Space Weather, National Space Science Center, Chinese Academy of Sciences, Beijing, 100190, China}
  \address[id=aff3]{School of Astronomy and Space Sciences, University of Chinese Academy of Sciences, Beijing, 100049, China}

   \runningauthor{Xu et al.}
   \runningtitle{Correction for the Weakening of Magnetic Field}

   \date{Received~~; accepted~~}

\begin{abstract}
The magnetic field inside the sunspot umbra, as observed by the Full-disk MagnetoGraph (FMG) onboard the Advanced Space based Solar Observatory (ASO-S), was found to be experiencing a weakening. To address this issue, we employed a method developed by \cite{Xu2021} to correct the weakening in the data of 20 active regions observed by FMG during the period spanning December 29, 2022, to July 23, 2023. Research has revealed that the onset of magnetic field weakening occurs at a minimum magnetic field strength of 705 G, with the peak strength reaching up to 1931 G. We computed the change ratio ($R_{1}$) of the unsigned magnetic flux within the sunspot umbra, considering measurements both before and after correction. The change ratio ($R_{1}$) spans from 26\% to 124\%, indicating a significant increase in the unsigned magnetic flux within sunspot umbrae observed by FMG after correction. To illustrate this, we selected four active regions for comparison with data from the Helioseismic and Magnetic Imager (HMI). After correction, it is found that the unsigned magnetic flux in sunspot umbrae measured by FMG aligns more closely with that of HMI. This supports the effectiveness of the corrective method for FMG, despite imperfections, particularly at the umbra-penumbra boundary.

\end{abstract}


\end{opening}

\section{Introduction}           
\label{sect:intro}

The strong magnetic field of sunspots has been the focus of exploration for over a century, dating back to its initial discovery by \cite{Hale1908}. A local relationship between the continuum intensity and magnetic field (hereafter denoted as $I-B$) in sunspot umbrae was established: the smallest intensity always corresponds to the largest magnetic field (e.g., \citealt{Martinez1993, Norton2004}). \cite{Leonard2008} studied the $I-B$ relationship of sunspots using 272 samples observed at the San Fernando Observatory and the National Solar Observatory, Kitt Peak. They found that the $I-B$ relationship is not linear beyond a field strength of about 700 G.  \cite{Moon2007} found that the absolute value of the Michelson Doppler Imager (MDI) flux density had a negative correlation with the MDI intensity for a high-intensity area larger than about 900, for the remaining area, both quantities had positive correlations, which were caused by the Zeeman saturation effect. \cite{Ulrich2002} provided a comprehensive analysis of the underlying causes for saturation effects encountered in the Mount Wilson 150-foot tower telescope system. This instrument utilized the spectral line Fe {\sc i} $\lambda$ 5250 {\AA} (Land\'{e} factor {\it g}=3) for its observations. They pointed out that most spectral lines used for magnetic measurements are subject to the saturation effect for at least some parts of their profile. \cite{Sakurai1995} emphasized that the large Land\'{e} factor of Fe {\sc i} 5250 {\AA} line can result in strong magnetic signal saturation at high field strengths, while the Fe {\sc i} 6302.5 {\AA} line, characterized by a Land\'{e} factor {\it g}=2.5, is less susceptible to this effect. Nevertheless, despite utilizing the Fe {\sc i} 6302.5 {\AA} line, the Solar Flare Telescope at Mitaka, Japan, still observed a notable decrease in magnetic signals at high field strengths. This finding implies that the Fe {\sc i} 6302.5 {\AA} line might not be as resistant to magnetic saturation effects as was previously believed. 

The saturation or weakening of magnetic signals at strong field strengths is more severe in filter-type magnetographs compared to spectral-type magnetographs. It is well established that modern magnetographs primarily measure solar polarized light, which can be described using the four Stokes parameters ($I$, $Q$, $U$ and $V$). The Stokes $Q$ and $U$ parameters represent the linear polarization related to the transverse magnetic field component, while the Stokes $V$ parameter represents the circular polarization corresponding to the line-of-sight (LOS) magnetic field component. In this study, we are primarily interested in the LOS component. Filter-type magnetographs observe circular polarization at a single wavelength and calibrate the LOS magnetic field from images of Stokes $V$ under the assumption of a weak magnetic field, where the relationship between LOS field strength and the Stokes parameter $V/I$ is approximately linear. In contrast, spectral-type magnetographs can obtain the spectral profiles of the Stokes parameters and invert for the vector magnetic field by employing the polarized radiative transfer equations within a specific atmospheric model. The Solar Optical Telescope (SOT) on  Hinode (\citealt{Kosugi2007}) comprises two distinct types of polarimeters: the Spectro-polarimeter (SP) obtains Stokes parameters spectral profiles of  Fe {\sc i} 6301.5 {\AA} (Land\'{e} factor {\it g}=1.667)/6302.5 {\AA} lines and the  Narrowband Filter Imager (NFI) obtains Stokes $I$ and $V$ images at a single-wavelength of Fe {\sc i} 6302.5 {\AA}. \cite{Chae2007} discovered a decrease in Stokes $V/I$ in sunspot umbral regions observed through the NFI. To rectify this, they conducted cross-calibration between the NFI's Stokes $V/I$ measurements and the LOS magnetic field data obtained by the SP. Meanwhile, \cite{Moon2007} utilized concurrent MDI intensity and magnetogram data. They employed a comparison between the SP and MDI flux densities to address saturation issue in the magnetic field measurements obtained by MDI. The Solar Magnetic Field Telescope (SMFT) located at the Huairou Solar Observing Station (HSOS) is a filter-type magnetograph that observes Stokes $V$ at $-0.075$ {\AA} off the spectral line center of Fe {\sc i} 5324.19 {\AA} (\citealt{Ai1986}). The Land\'{e} factor {\it g}=1.5 and the equivalent width of the line is 0.33 {\AA}. It was discovered in SMFT data that the LOS magnetic field weakens in sunspot umbrae. \cite{Plotnikov2021} conducted a cross-calibration between SMFT data and magnetograms provided by the Helioseismic and Magnetic Imager (HMI: \citealt{Schou2012}), which acquires a spectral profile with six wavelength points. They introduced a non-linear relationship between the Stokes $V/I$ and LOS magnetic field, aiming to eliminate the observed weakening of magnetic field inside sunspot umbra. To address this magnetic field weakening issue, \cite{Xu2021} developed a method based on the correlation between the Stokes $V/I$  and $I/I_{m}$ ($I_{m}$ represents the maximum value of Stokes $I$) observed by SMFT. This approach relies solely on data from a single instrument, thereby avoiding systematic errors associated with cross-comparisons.

The Full-disk MagnetoGraph (FMG) is a payload onboard the Advanced Space based Solar Observatory (ASO-S: \citealt{Gan+etal+2019}) designed to measure solar photospheric magnetic fields (\citealt{Deng+etal+2019}). Routine observations with the FMG are conducted at a single wavelength position of the Fe {\sc i} 5324.19 {\AA} line, utilizing a linear calibration under the weak-field approximation (\citealt{Su+etal+2019, Liu2023}). Therefore, it is not surprising to discover a decrease in the LOS magnetic field in sunspot umbrae. To rectify this issue and enhance the quality of magnetic field data from the FMG, we will employ the method developed by \cite{Xu2021}. This paper is organized as follows: Section 2 provides a description of our methodology and data reduction procedures. Section 3 presents the primary results. Finally, Section 4 summarizes our main conclusions and offers further discussions.

\section{Method and Observations}
\label{sect1}

\cite{Xu2021} developed a method that can automatically identify the threshold of saturation in LOS magnetograms observed by SMFT and correct it. This method relies on the relationship between Stokes $V/I$ and $I/I_{m}$ to estimate the saturation threshold and reconstruct Stokes $V/I$ in strong magnetic field regions to correct for saturation. $I_{m}$ represents the maximum value of Stokes $I$ for the whole intensity image of an active region, which varies with the active region. The effectiveness of the algorithm  was demonstrated by comparing it with magnetograms obtained by HMI and a sample of 175 active regions observed by SMFT. In this study, we adopt this method to address the weakening of magnetic fields in sunspot umbrae observed by FMG. To illustrate the effectiveness of correction, we introduce two parameters:

\begin{equation}  \label{Eq-R1}
 R_{1} =  \frac{F_{\rm af}-F_{\rm bf}}
                 {F_{\rm bf}} \times 100\% \,,
\end{equation}
and

\begin{equation}  \label{Eq-R2}
 R_{2} =  \frac{|F_{\rm h}-F_{\rm af}|}
                 {F_{\rm h}} \times 100\% \,,
\end{equation}
where $F_{\rm bf}$ and $F_{\rm af}$ represent the unsigned magnetic flux inside the sunspot umbra observed by FMG before and after correction, respectively, and $F_{\rm h}$ represents the unsigned magnetic flux in the same region as observed by HMI. Therefore, $R_{1}$ represents the change ratio of unsigned magnetic flux after and before correcting for the weakening of magnetic field in sunspot umbrae observed by FMG, while $R_{2}$ represents the ratio of unsigned magnetic flux differences between HMI and FMG after correction. For comparison, we also introduce $R_{3}$ which represents the ratio of unsigned magnetic flux differences between HMI and FMG before correction.

\begin{equation}  \label{Eq-R3}
 R_{3} =  \frac{F_{\rm h}-F_{\rm bf}}
                 {F_{\rm h}} \times 100\% \,.
\end{equation}

We selected 20 active regions observed by FMG from December 29, 2022 to July 23, 2023. Four of these active regions have been utilized in \cite{Xu2024}  and the same data reduction method was employed. The remaining 16 active regions were chosen when they were closest to the solar disk center and had the smallest satellite orbit velocity.

\begin{figure}
\centerline{\includegraphics[width=0.99\textwidth,clip=]{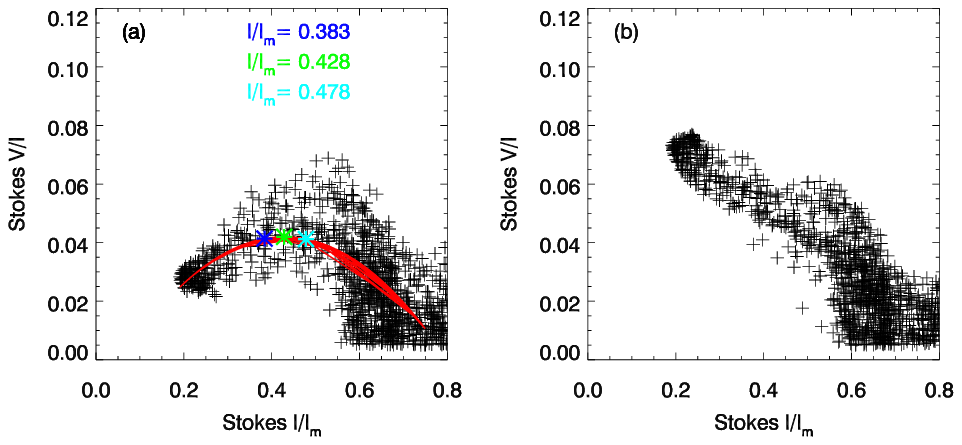}}
\vspace{0.01\textwidth}     
\centerline{\includegraphics[width=0.99\textwidth,clip=]{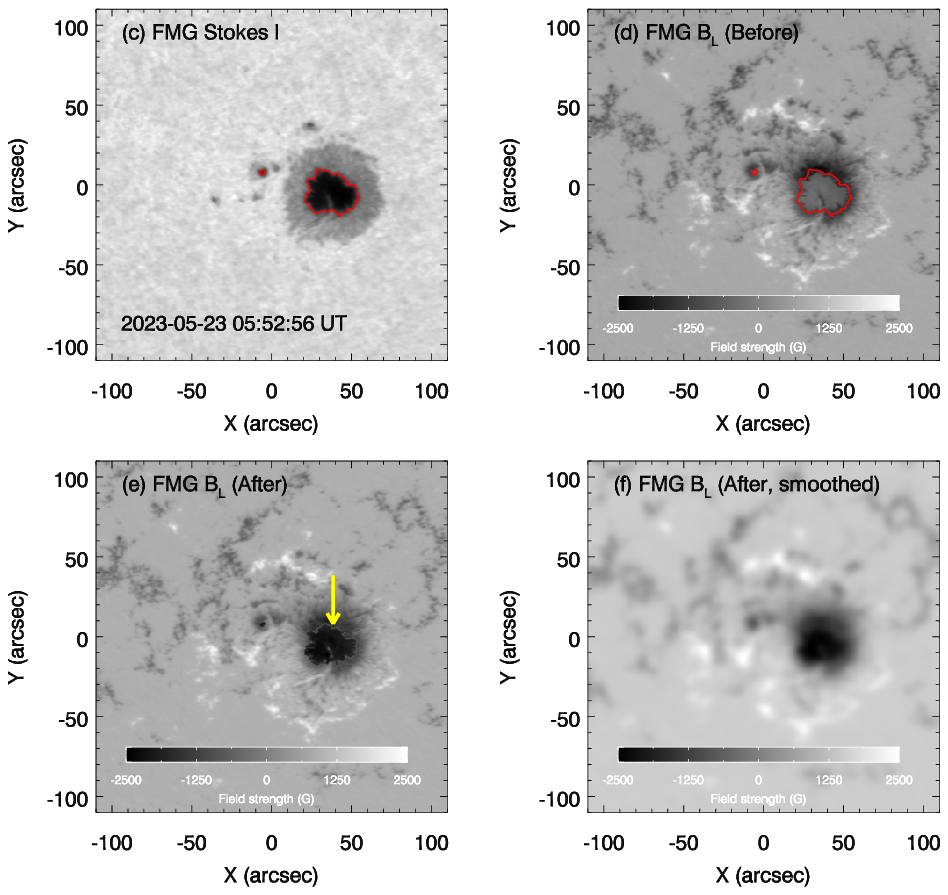}}
\vspace{0.01\textwidth}
\caption{Panel (a) displays the $I/I_{m}-|V/I|$  relationship of  NOAA 13310 observed by FMG. The green asterisk presents the apex of the second-order polynomial fit (red line). The cyan and blue asterisks present the $\pm 1\sigma$ uncertainty of apex. The corresponding $I/I_{m}$ values are marked using the same color as asterisks. Panel (b) displays the $I/I_{m}-|V/I|$  relationship after correcting for the weakening magnetic field  in sunspot umbra. Panel (c) is the map of Stokes $I$ and panel (d) shows the LOS magntogram before correction. The red contour represent $I/I_{m}$=0.428. Panels (e) and (f) show the corrected LOS magnetograms without and with smoothing the data, respectively.}\label{eg}
\end{figure}

\section{Results}

\begin{figure}
\centerline{\includegraphics[width=1.0\textwidth,clip=]{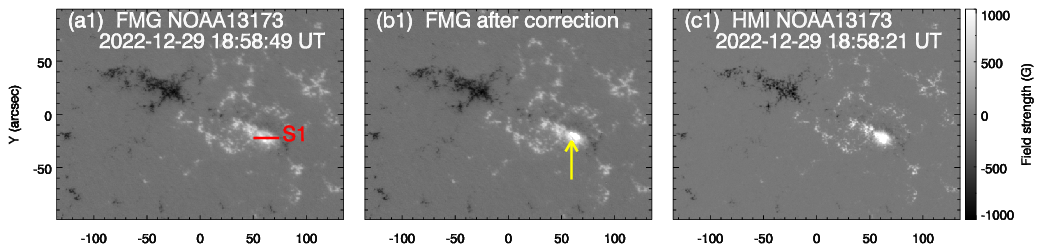}}
\vspace{-0.005\textwidth}     
\centerline{\includegraphics[width=1.0\textwidth,clip=]{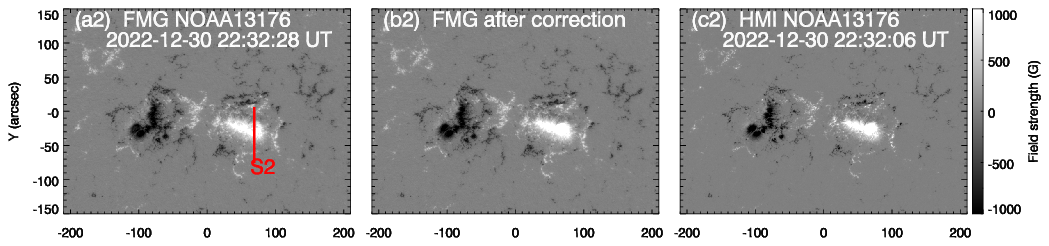}}
\vspace{-0.005\textwidth}
\centerline{\includegraphics[width=1.0\textwidth,clip=]{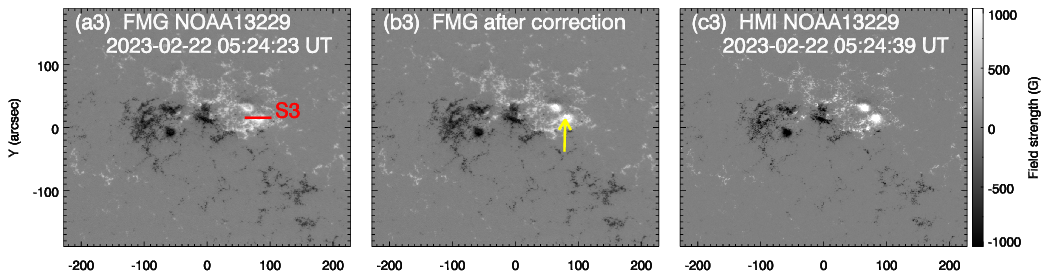}}
\vspace{-0.005\textwidth}     
\centerline{\includegraphics[width=1.0\textwidth,clip=]{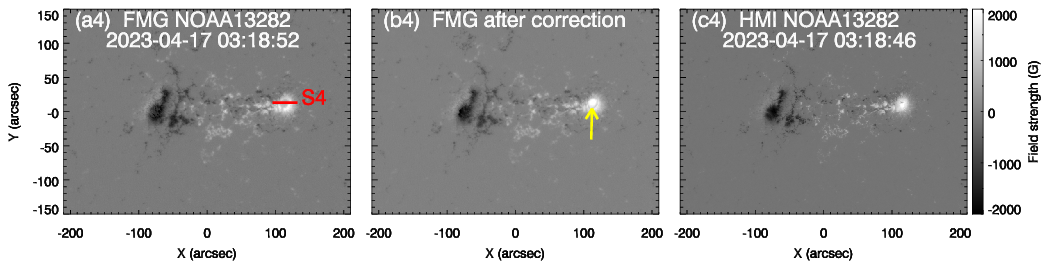}}   
\caption{Panels (a1)--(a4) display the original LOS magnetograms of four active regions observed by FMG. Panels (b1)--(b4) represent the LOS magnetograms after correcting for magnetic field weakening in sunspot umbrae observed by FMG. Panels (c1)--(c4) show the LOS magnetograms observed by HMI.}\label{mag}
\end{figure}

\subsection{Test of the Method}

We take active region NOAA 13310 acquired by FMG as an example to test the method. The relationship between the values of $I/I_{m}$ and $|V/I|$  is given in Figure \ref{eg}(a), which is similar to Figure 2 in \cite{Xu2021}. It is found that the second-order polynomial (red line) gives a well fit when $|V/I|$ $>$ 0.01 and $I/I_{m}$ $\leq$ 0.75. The $I/I_{m}$  corresponding to the apex (green asterisk) is 0.428. $I/I_{m}$ and $|V/I|$ have a negative correlation for area $I/I_{m} > 0.428$, and a positive correlation for area $I/I_{m}< 0.428$ which is  inside the red contour in Figure \ref{eg}(c). The weakening magnetic field  is found for the area inside the red contour in Figure \ref{eg}(d), owing to the linear calibration is adopted by FMG. The $I/I_{m}$ corresponding to the apex can be as the threshold (hereafter denoted as $I_{0}$) for occurrence of magnetic weakening. We reconstruct Stokes $V/I$ for pixels where $I/I_{m}$ $<$ $I_{0}$ using Equation \ref{Eq-V} (Equation 3 in \citealt{Xu2021}). 
\begin{equation}  \label{Eq-V}
{V \over I}^s =\left|{a_{2} \over a_{1}}\left(\left|{V \over I}\right|-c_{1}\right)+ c_{2}\right|\cdot sign\left({V \over I}\right) \,, \\
\end{equation}
where ($a_{1}$, $c_{1}$) and ($a_{2}$, $c_{2}$) are linear fitting coefficients corresponding to area $I/I_{m}$ $<$ $I_{0}$ and $I/I_{m}$ $\geq$ $I_{0}$, respectively. $a_{1}$ and $a_{2}$ are slopes, $c_{1}$ and $c_{2}$ are constants. After re-calculating, the relationship between the values of $I/I_{m}$ and $|V/I|$ is given in Figure \ref{eg}(b), and a negative correlation is found.

The FMG LOS magnetic field $B_{L}$ can be re-calibrated from Equations \ref{Eq-V1} and  \ref{Eq-V2}:

\begin{equation}  \label{Eq-V1} 
 B_{L}= C_{L}{V \over I}^s \,, \ \ \ \ \ \ {I \over I_{m}} < I_{0} \,, 
\end{equation}
and
\begin{equation}\label{Eq-V2}
 B_{L}=C_{L}{V \over I}  \,, \ \ \ \ \ \ \  {I \over I_{m}}  \geq  I_{0} \,, 
 \end{equation}
where $C_{L}$ is calibration coefficient. FMG adopts 30000 G for routine calibration. The re-calibrated $B_{L}$  map is shown in Figure \ref{eg}(e). It can be seen that the weakening magnetic field  in sunspot umbra has been corrected. But the discontinuity (yellow arrow) arises at the umbra--penumbra boundary. We applied a Gauss-smooth function to smooth the corrected data by 4 pixels. The smoothed $B_{L}$  map is displayed in Figure \ref{eg}(f). It can be seen that the discontinuity has been eliminated, however, the spatial resolution has also been reduced. 

\begin{table}
\caption[]{The information regarding the four active regions that were simultaneously observed by FMG and HMI}\label{4ar}
 \begin{tabular}{ccccccccc}
  \hline
   NOAA    &  Location  & Date & Time (UT)        & B$_{\rm s}$ (G)\tabnote{The magnetic field strength at the point where it starts to decrease.}    & $R_{1}$ (\%)\tabnote{The change ratio of unsigned magnetic flux after and before correcting the weakening of magnetic field in sunspot umbrae observed by FMG.}   & $R_{2}$ (\%)\tabnote{The ratio of unsigned magnetic flux differences between HMI and FMG after correction.} & $R_{3}$ (\%)\tabnote{The ratio of unsigned magnetic flux differences between HMI and FMG before correction.}   \\  
  \hline
13173 & N24W15    & 2022-12-29   &18:58:49    & 958      & 60    &  5     &  41      \\  
13176 & N19E12     & 2022-12-30   &22:32:28    & 1015    & 57    &  5    &  33      \\  
13229 & N25W05    & 2023-02-22   &05:24:23    & 705      & 44    &  19   & 43     \\ 
13282 & N11E10     & 2023-04-17   &03:18:52    & 1071    &  69   &  16  & 30    \\  
  \hline
\end{tabular}
\end{table} 

\subsection{Correction for FMG Data}

\begin{figure}
\centerline{\includegraphics[width=1.0\textwidth,clip=]{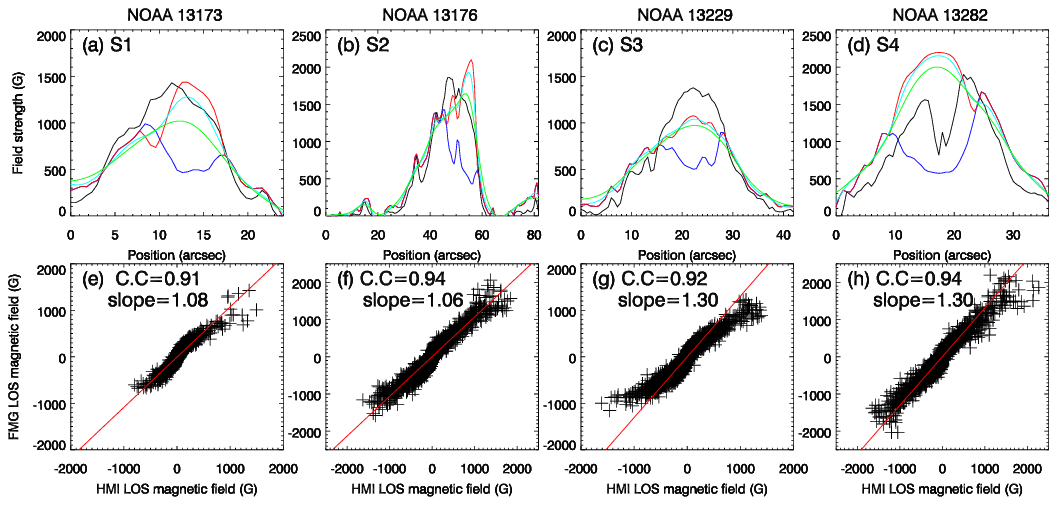}}
\caption{Panels (a)--(d)  are magnetic field distribution along slices S1--S4 as indicated in Figure \ref{mag}. The black and blue solid lines in each panel represent HMI and FMG, repectively. The red, cyan and green solid lines represent the FMG magnetic field after correcting for weakening of the magnetic field in sunspot umbra. The red line indicates that the corrected magnetic field is not smoothed, while the cyan and green lines indicate that the corrected magnetic field has been smoothed by 2 and 4 pixels. Panels (e)--(h) feature the scatter plots of magnetic field (FMG vs. HMI) after correcting for the weakening of magnetic field in sunspot umbra observed by FMG, no smoothing applied.  C.C is the correalation coefficient of magnetic field.
}\label{sation}
\end{figure}

Figures \ref{mag}(a1)--(a4) present the original magnetograms of four active regions observed by FMG. Within the umbrae, a decrease in magnetic field strength is observed. Figures \ref{mag}(b1)--(b4) display the magnetograms after correcting for magnetic field weakening in sunspot umbrae. It is evident that the magnetic field strength increases within sunspot umbrae, aligning more closely with the HMI-45s magnetograms in Figures \ref{mag}(c1)--(c4). However, a discontinuity indicated by yellow arrows at the boundary of umbrae and penumbrae is visible. To conduct a detailed analysis, four slices passing through the sunspots were created, as indicated by red lines in Figures \ref{mag}(a1)--(a4). The distributions of magnetic field along slices S1-S4 are depicted in Figures \ref{sation}(a)--(d). It is observed that the magnetic field strength in sunspot umbrae significantly increases after correction, as indicated by comparing the red and blue lines. Specifically for slices S2 and S3, the magnetic fine structures of FMG align well with those of HMI (black lines). However, we also identify discontinuities or errors at the positions where the magnetic field starts to decrease. To eliminate this discontinuity, we applied a Gaussian smoothing function to the corrected data. The cyan and green lines in Figures \ref{sation}(a)--(d) indicate that the corrected magnetic field has been smoothed by 2 and 4 pixels, respectively. It is observed that the discontinuity has been eliminated, but the magnetic field outside the umbrae was also altered. It is evident that the more data is smoothed, the more effective the elimination of discontinuities. The scatter plots of magnetic field (FMG vs. HMI) after correcting for magnetic field weakening in sunspot umbra observed by FMG are presented in Figures \ref{sation}(e)--(h). The relationship is approximately linear. We employed a linear fitting to the data with errors included. The error is 10.2 G for HMI 45-s magnetograms (\citealt{Liu2012}) and 15 G for FMG in normal mode (\citealt{Deng+etal+2019}). The correlation coefficients and slopes are all slightly higher compared to those reported in \cite{Xu2024}. 

Additionally, we calculated the magnetic field strength $B_{s}$ at the point where it starts to decrease, as well as $R_{1}$, $R_{2}$ and $R_{3}$. The results are summarized in Table \ref{4ar}. The lowest $B_{s}$ value is 705 G, while the highest is 1071 G. $R_{1}$ ranged from 44\% to 69\%, indicating that the unsigned magnetic flux in sunspot umbrae observed by FMG increased by 44\%--69\% after correction. Furthermore, $R_{2}$ was significantly smaller than $R_{3}$, indicating that the magnetic flux in sunspot umbrae observed by FMG aligns more closely with that of HMI after correction.

For the other 16 active regions, we make the same analysis and find a similar result. Table \ref{16ar} gives the information and results of the 16 active regions. The lowest $B_{s}$ is 991 G and the highest is 1931 G. The lowest $R_{1}$ is 26\%  and the highest is 124\%. $B_{s}$ and $R_{1}$ are different for different active regions, but we do not find any relationship between $B_{s}$ and $R_{1}$. 

\begin{figure}
\centerline{\includegraphics[width=1.0\textwidth,clip=]{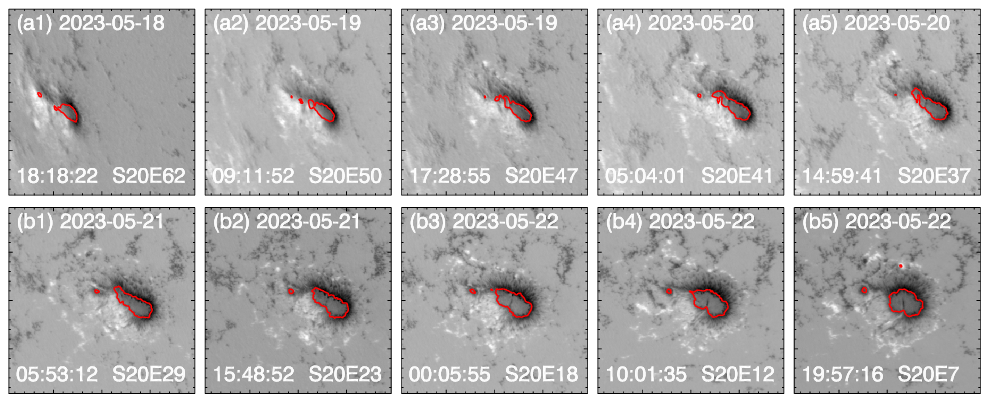}}
\vspace{-0.015\textwidth} 
\centerline{\includegraphics[width=1.0\textwidth,clip=]{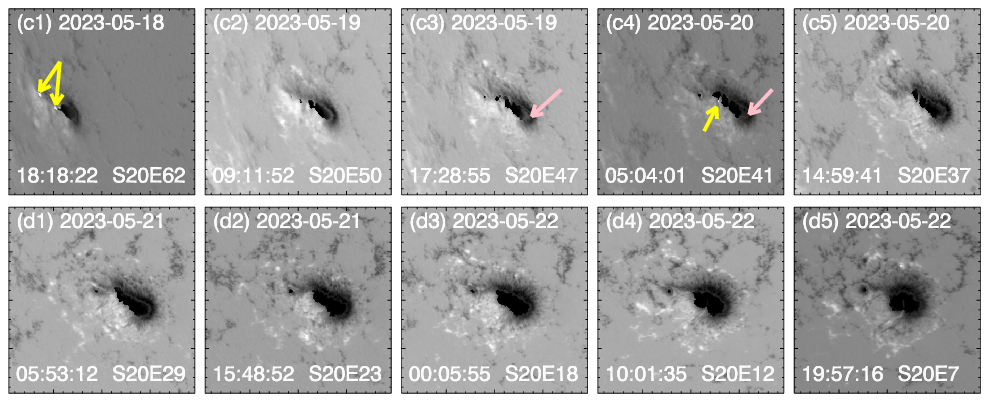}}
\caption{Panels (a1)--(b5) display the original LOS magnetograms of NOAA 13310 observed by FMG from May 18 to 22, 2023. The weakening magnetic field persists inside the red contours.  Panels (c1)--(d5) exhibit the corrected LOS magnetograms of the same active region. The yellow arrows mark the abnormal points. The pink arrows point out the uncorrected area. The size of magnetograms is $192.5^{\prime\prime} \times 192.5^{\prime\prime} $.
}\label{muti}
\end{figure}

\begin{figure}
\includegraphics[width=1.0\textwidth,clip=]{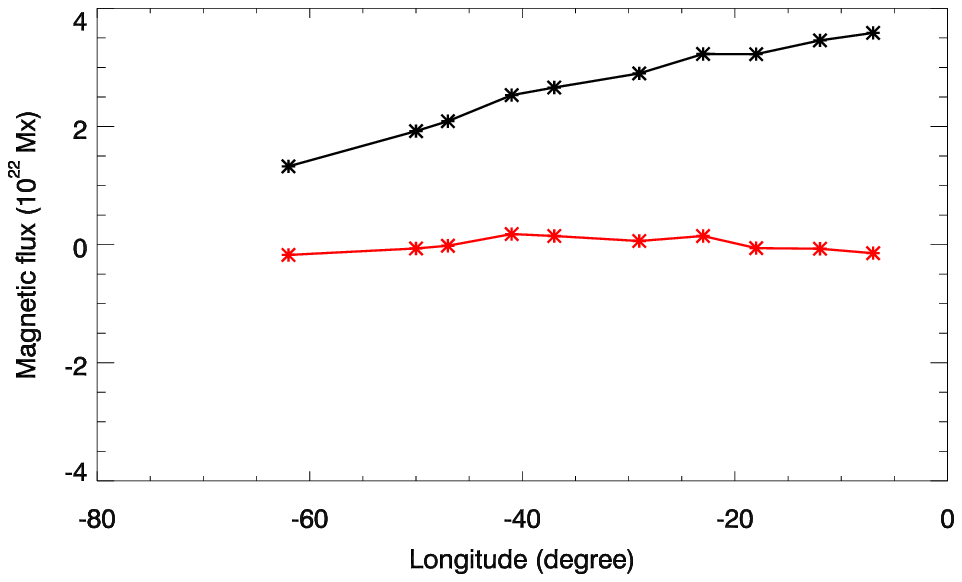}
\caption{The variation of magnetic flux with longitude of  NOAA 13310 from May 18 to 22, 2023. The black and red lines represent the unsigned magnetic flux and the  residual flux, repectively.}\label{flux}
\end{figure}

To investigate the effect of location on the correction algorithm, we selected a relatively stable active region NOAA 13310 and traced it from east to the solar disk center (from May 18 to 22, 2023). We used the FMG data when the satellite orbit velocity was near zero to diminish its influence on magnetic field strength. From Figures \ref{muti}(a1)--(b5), one can see that the weakening of magnetic field persists for all magnetograms. The red contours represent $I/I_{m}$ values calculated by the algorithm, indicating the occurrence of magnetic weakening. It is found that the algorithm is independent on solar disk location when detecting the weakening magnetic field. Figures \ref{muti}(c1)--(d5) show the magnetograms after correcting for the weakening magnetic field. Some abnormal points (yellow arrows) are found  and the magnetic field in certain areas have not been fully corrected (pink arrows) when the active region is far from disk center. The correction is more accurate when the active region is close to the disk center. We calculated the unsigned magnetic flux from the corrected magnetograms. From Figure \ref{flux}, one can see that the unsigned flux gradually increases from east toward the solar disk center (black line). We employed a linear fitting to the data and then subtracted the fitting result from the unsigned flux to remove the general trend. There is no systematic variation in the residual flux (red line). 

\begin{table}
\caption[]{The information regarding the 16 active regions observed by FMG}\label{16ar}
 \begin{tabular}{ccccccccc}
  \hline
   NOAA    &  Location  & Date & Time (UT)        & B$_{\rm s}$ (G)    &$R_{1}$ (\%)     \\  
  \hline
13273 & N10W11    & 2022-12-06   &02:36:55    & 1532   & 35      \\  
13285 & S17E04     & 2023-04-26   &17:36:15    &  1626  & 58      \\
13288 & S22E10     & 2023-04-26   &17:36:15    &  1135  & 26     \\ 
13293 & N13W07    & 2023-05-06   &12:36:39   & 1769   & 36     \\
13294 & N08E03     & 2023-05-07   &20:48:44   &  1931  &  66       \\ 
13296 & N08E03     & 2023-05-07   &03:30:10    & 1696   & 27    \\
13297 & N08W04    & 2023-05-08   &01:46:34    & 1616   & 32    \\ 
13310 & S20E02     & 2023-05-23   &05:52:56    & 1257   & 96     \\ 
13314 & N15W04    & 2023-05-23   &05:52:56    &  1817  & 124     \\
13315 & S16E05      & 2023-05-26   &19:56:44    &  1259  & 34     \\
13321 & S15W01     & 2023-06-04   &00:53:01    &  1347  & 99     \\
13339 & S18E01     & 2023-06-22   &09:06:27     &   991   & 76    \\
13340 & S21E05     & 2023-06-22   &09:06:27     &   1855 & 27   \\
13354 & S15W03     & 2023-06-28   &02:28:02    &  1557  & 48     \\
13373 & N07W01     & 2023-07-19   &01:34:25    &  1533  & 68    \\
13377 & S08W07     & 2023-07-23   &03:13:18    &  1486  &  80   \\
  \hline
\end{tabular}
\end{table}

\section{Conclusions and Discussions}

We conducted a study on the weakening magnetic field within the sunspot umbrae of 20 active regions observed by FMG. It has been demonstrated that the method developed by Xu et al. (2021) can be directly applied to FMG data, with only minor modifications required for the range of Stokes $V/I$ and $I/I_{m}$ when performing polynomial fits. The thresholds for magnetic field weakening to occur range from 705 G to 1931 G. After correcting for the magnetic field weakening, the unsigned magnetic flux within sunspot umbrae observed by FMG increased by 26\%--124\%, and the difference ratio of unsigned magnetic flux between HMI and FMG decreased from 30\%--43\% to 5\%--19\%.

Despite its effectiveness, this method possesses certain limitations. While smoothing the data eliminates the discontinuity at the boundary between umbra and penumbra, it also diminishes spatial resolution, thus compromising the recognition of fine structures. The method relies on the relationship between $V/I$ and $I/I_{m}$ which is an extension of the $I-B$ relationship that varies across different umbrae (\citealt{Norton2004}). Notably, the correction's accuracy diminishes when active regions are situated far from the disk center. Therefore, this method has not yet been integrated into the data production process, and its use should be determined based on the specific research context.

The Fe {\sc i} 5324.19 {\AA} spectral line is the most commonly used to observe magnetic fields at HSOS. A numerical calculation by \cite{Zhang2019} showed that the Stokes parameters of the spectral line have different sensitivities in the quiet Sun and sunspot atmosphere. The linear approximation for Stokes $V$ can only be used for relatively weak magnetic fields below 2000 G for the quiet Sun and 1000 G for the umbra. Our observations reveal that the magnetic field strength at which magnetic field weakening begins ranges from 705 G to 1931 G (average value $\sim$ 1400 G ). This suggests that the magnetic field range suitable for using linear approximation varies depending on the sunspot. The calibration process from polarized light to magnetic field is complex, and both numerical calculations and observations suggest that different calibration coefficients should be used for different regions of the Sun. Additional research is required for the calibration of filter-type magnetographs measuring circular polarization at a single-wavelength position.

\begin{acks}
We are sincerely grateful to the anonymous reviewer whose comments helped us to significantly improve the article. This work is supported by National Key R\&D Program of China (Grant No. 2022YFF0503800, 2022YFF0503001, 2021YFA1600500), National Natural Science Foundation of China (Grant No. 12273059, 11427901, 12173049, 12073040, 12003051, 12373057),  Beijing Natural Science Foundation (Grant No. 1222029), and  the Strategic Priority Research Program of the Chinese Academy of Sciences (Grant No. XDB0560000). ASO-S mission is supported by the Strategic Priority Research Program on Space Science, the Chinese Academy of Sciences, Grant No. XDA15320000. We acknowledge the free data usage policy of the SDO/HMI. The HMI data used in this paper is taken from \url{http://jsoc.stanford.edu}.

\end{acks}

\end{article}

\end{document}